\documentstyle[prl,aps,epsf,floats]{revtex}

\begin{document}
\draft

\twocolumn[\hsize\textwidth\columnwidth\hsize\csname @twocolumnfalse\endcsname
\title{  
Effects of gating and contact geometry
on current through conjugated molecules covalently
bonded to electrodes
} 
\author{ A.M. Bratkovsky and P.E. Kornilovitch }
\address{
Hewlett-Packard Laboratories, 1501 Page Mill Road, 1L, Palo Alto, 
California 94304
}

\date{ April 27, 2002 }
\maketitle

\begin{abstract}

We study the effects of gating and contact geometry
on current through self-assembled monolayers of conjugated molecules
strongly coupled to gold  electrodes by sulfur ``anchor groups''.
The current changes by more than an order of magnitude
depending on the  angle between the axis of the benzene-dithiolate
molecules and the
normal to the electrode on the less coordinated ``top site''
position. The effect  
of gating is also much stronger in this case compared to higher coordinated
``hollow site'' binding of the molecule on a Au(111) surface. 
The large hybridization
of the molecular states with electrode states for the hollow site leads to 
practically ohmic current-voltage characteristics. 
Changes in molecule-electrode geometry accompanying the gating of the
SAM may be the reason for strong changes of the conductance observed
by  Sch\"on {\it et al.} in the ``slot'' geometry.

\end{abstract}

\vskip2pc] \narrowtext


\section{Introduction}

Studies of electron transport through organic molecules (molecular films),
viewed as the possible components of molecular electronic devices, are a
very active area of research \cite{Aviram,Aviram98,RatnerRev99}. Although
the rectifying properties of the molecules in two-terminal devices were
demonstrated in 1990s\cite{recti93}, only recently have {\em three}-terminal
devices based on a self-assembled monolayer (SAM) of simple conjugated
molecules containing either one or two phenyl rings C$_{6}$H$_{6}$ or two or
three thiophen rings C$_{4}$H$_{4}$S been fabricated\cite{SchonSAM1}. The
molecules were gated from the edge of the vertical device through
$L=$30 nm of SiO$_{2}$ gate oxide (30 nm was the nominal oxide
thickness, but the
high-resolution transmission electron microscopy
showed much thinner oxide barrier of 4-5nm \cite{SchonBDT01}).
In the slot geometry of this experiment, the drain
current $I_{d}$ through only several thousand molecules close to the gate
oxide should have been affected by the gate voltage $V_{g}$. The measured
conductance $dI_{d}/dV$ at zero source-drain bias was about 5 $\mu $%
S/molecule. The gating field on the molecules should be small because the
length of the molecules (the ``channel'' length) is tiny, $2t=1-2$ nm, so
that the geometrical aperture factor is small, $t/L\ll 1.$ In spite of this,
Sch\"{o}n {\it et al.} have initially estimated a dramatic change of
conductance of the 
molecules by a factor to be about $10^{7}$ (it was later determined that the
effect was overestimated, probably because of disorder and other
factors in pure SAMs\cite{SchonBDT01}), with a large transconductance $%
dI_{d}/dV_{g}=12-13$ mA/V for pure SAMs of biphenyl molecules
and about 6 mA/V for phenyls\cite{SchonSAM1} at room
temperature. {\em Dilute} systems of conducting molecules
(namely, 4,4'-biphenyldithiol and some other species) embedded in a matrix of
non-conducting alkanedithiols have also been measured with the aim to sample
individual conducting molecules\cite{SchonSAM2}. Sch\"on {\em et al.}
have found that the 
conductance through the molecule is about 0.1$\mu $A/V at zero drain voltage 
$V=0$, and approaches one conductance quantum $2e^{2}/h=77.5\mu $A/V at
relatively small gate voltage $V_{g}\approx -0.3$ V. The histograms of
conductance measured on a series of devices suggested that the conductance
through the gated molecules was quantized in the fundamental units of $%
2e^{2}/h.$ The conductance peak position varied with the gate voltage $V_{g}$
from device to device. The transconductance $dI_{d}/dV_{g}$ was found to be
in the range of $150\mu $A/V at 4K, {\em decreasing} to about 10$\mu A/V$ at
room temperature (in the earlier measurements the peak value of $%
dI_{d}/dV_{g}$ was estimated of about 13mA/V \cite{SchonSAM1} for the
biphenyl molecules). The corresponding modulation of conductance in the film
of individual dithiol molecules has been estimated to be much
larger than $10^{3}$. This is a huge effect indeed if it is to be
explained by a mechanism related to the behavior of individual molecules.

Very recently this work has been extended to SAMs of 
benzene-(1,4)-dithiolate (-S-C$_{6}$H$_{4}$-S-), the simplest
conjugated molecules, referred to below
as BDT, with only one phenyl ring\cite{SchonBDT01}. 
The area of the measured device comprised about $%
10^{4}$ molecules, mostly non-conducting alkanes with a small
fraction (10$^{-4})$ of the conducting BDT molecules\cite{SchonBDT01}. In
this geometry, only one or a few BDT molecules should have contributed to
transport. The first peak in BDT conductance has been  found at a bias of
about 2.1V with the corresponding value of 25$\mu$A/V \cite{SchonBDT01}.
Interestingly, the conductance
appeared to be much larger than that found in the earlier
break-junction experiments 
by Reed {\it et al.}\cite{Reed97} who observed the first peak  in
conductance of 0.05$\mu $A/V at 1.4V. 
Sch\"{o}n and Bao have also
observed a jump in conductance from 0.7-1.0 $\mu $A/V to much
larger values close to the conductance quantum $77.5$ $\mu $A/V at the gate
voltage $V_{g}\approx -0.4$ V and the drain voltage of about $-0.7$ V (this
is a jump of almost two orders of magnitude). The position of this
switching shows some {\em %
hysteresis} when the gate voltage is varied in the range $-0.6<V_{g}<0$ V.
This hysteretic behavior cannot possibly come from the individual molecule
behavior and should be related to some {\em extrinsic} mechanism, as we
shall discuss below. Also interesting is the observed change in the fine
structure of the conductance: there is a series of small peaks 45 meV apart
from each other on the conductance curve for $V_{g}=0,$ and the period
shrinks down to 40 meV at $V_{g}=-0.5$ V. Such peaks are usually attributed
to tunneling assited by molecular vibronic excitations, so the change in the period
indicates a possible change in molecular conformation. These
abrupt changes cannot be explained by the electric field created by
the gate, since the corresponding field on the molecule should be small.

With regards to a possible origin of the observed behavior, we first mention
the importance of the geometry of the molecule-electrode contact\cite{KB01}. 
If the orientation of the molecule with respect to an
electrode changes, so does its conductance. Indeed, we have found earlier
that the conductance of BDT (or any other conjugated thiol-terminated
molecule) strongly depends on the angle $\theta $ between the molecular
``backbone'' and the normal to the gold surface
(Fig.~\ref{fig:site}). In a simple ``toy'' model, 
the current dependence is $\propto \sin ^{2}\theta $ in the regime of strong
resonant tunneling (large bias) and even stronger, $\propto \sin ^{4}\theta $ in the
regime of non-resonant tunneling (small bias)\cite{KB01}. Note that
the angular dependence of current is much stronger when
the end sulfur is in the less-coordinated ``top-site'' position above a
surface Au atom, compared to a ``hollow-site'' position,
when it is bonded to three surface gold atoms, see
Fig.~\ref{fig:site}. These bonding 
positions were considered in the literature as being the most favorable\cite
{Sellers}. Indeed, it has been provisionally
suggested that the 
conformational changes of the BDT with respect to electrodes, discussed in 
\cite{KB01}, are responsible for the observed ``switching'' behavior of
SAMFETs\cite{SchonBDT01}. Also, one cannot exclude that the observed changes
in conductance through BDT are caused by charge trapping-detrapping
processes close to the interface between the molecular layer and the
gate SiO$_{2}$.
\begin{figure}[t]
\epsfxsize=7cm
\epsffile{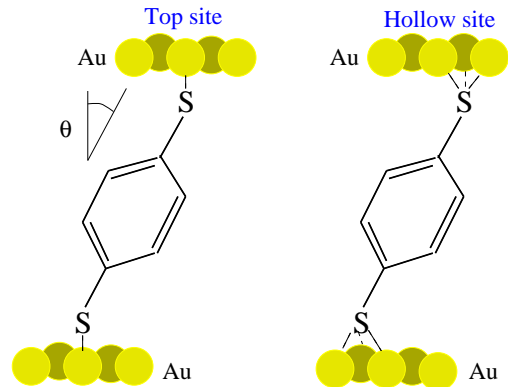}
\caption{
Schematic of the benzene-dithiolate molecule on top and hollow sites.
End sulfur atoms are bonded to one and 
three surface gold atoms, respectively.  $\theta$  is the tilting
angle. 
}
\label{fig:site}
\end{figure}

\section{Self-consistent calculations of electronic structure and transport}

In order to gain more insight into the origin of the observed extraordinary
dependence of molecular conductance on a weak gating field observed in Refs. 
\cite{SchonBDT01,SchonSAM2}, we have performed a series of self-consistent
calculations for different attachments of the BDT\ molecule to Au(111)
surface. We have found that the effects of charge redistribution 
as a function of molecular configuration and/or external field are
important, since the mismatch between the work functions of Au and BDT\
results in a considerable charge flow between the molecule and gold
electrodes. Without external bias, the lowest unoccupied molecular orbital (LUMO)
lies closer to the Fermi level of Au compared to the highest
occupied molecular orbital (HOMO). The current
through the molecule strongly depends on both the tilting angle
and the self-consistent charge
redistribution across the molecule. Although it is not clear {\em a priori}
what the sign of the combined effect would be,
we have found that our previous conclusion about strong orientation
dependence of the conductance through anisotropic conjugated $\pi -$%
orbitals and the s-orbitals on electrode Au remains valid\cite{KB01}.

BDT attaches strongly to the gold substrate by thiolate end groups -S- that
form covalent bonds with Au \cite{Reed97,Dorogi}. In order to 
properly account for such a bonding in the present calculations, 
the Au atom(s)
connected to S are treated separately from other gold atoms that
compose the gold electrode. The conductance is computed with the use
of the general procedure of Ref.~\cite{GF99}. The gold electrodes are described
by a tight binding model with nine $s-$, $p-$, and $d-$orbitals per each Au
atom with parameters from Ref.~\cite{Papa}. The equilibrium molecular
geometry is found by total-energy density functional minimization \cite
{Spartan}. The tight-binding parameters for the molecules and molecule-lead
interfaces are taken from the solid-state table of elements \cite{Harrison}.

The onsite energies in the present tight-binding model, which are very
important for finding the 
correct charge redistribution between the molecule and the electrodes,
have been estimated from the Hubbard model in the atomic 
limit. The energy of an isolated atom is approximated as $E_{m}=E_{0}-$ $\epsilon
\Delta q_{m}+\frac{1}{2}U\Delta q_{m}^{2}$. Here $E_{0}\ $is the energy of
a neutral atom with the atomic energy level at $-\epsilon <0\ $with respect
to the vacuum level (energy origin),  $\Delta q$ is  the excess charge
of an atom, and $U$ is the intraatomic Coulomb repulsion. In this
approximation, we obtain $\epsilon =\frac{1}{2}\left( A+I\right) ,$
$U=I-A$, where $A$ and $I$ are the experimental atomic values for the
affinity and the ionization energy, respectively.
These expressions have been used to estimate the following parameters
used in the  present work: 
$U=11.5,$ $\epsilon =7.8$ for H; $U=6.3,$ $\epsilon =5.2$ for
C; $U=7.8,$ $\epsilon =6.5$ for S; and $U=6.7,$ $\epsilon =5.9$ for Au (all
in units of eV).
We would like to mention that the use of different values for the
one-electron energies $\epsilon$, like the ones from
Ref.~\cite{Harrison}) leads to unphysically large charge transfers. 

We have calculated the current through BDT  on Au(111) in both the
top-site and the hollow-site positions. Including onsite and intersite
Coulomb interactions one finds that the onsite one-electron energies for
state $a$ at the site $m$ should be adjusted as

\begin{equation}
\epsilon _{ma}=\epsilon _{ma}^{0}+U_{m}\Delta q_{m}+\sum_{m^{\prime
}(m^{\prime }\neq m)}e\gamma _{mm^{\prime }}\Delta q_{m^{\prime }}+e\phi
_{m}^{I},  \label{eq:ema}
\end{equation}
where $\epsilon _{ma}^{0}$ are the onsite energies in a system with neutral
atoms, $\Delta q_{m}$ are the charges on sites, $\gamma _{mm^{\prime }}=1/|%
{\bf m}-{\bf m}^{\prime }|,$ $\phi _{m}^{I}$\ the image potential, and $e<0$
is the electron charge\cite{Kirczenow00}. The charge $\Delta q_{m}$ is found
self-consistently from the local density of states, which is given by the
site-projected imaginary part of the exact Green's function of the problem.
The total retarded Green's function $G_{mam^{\prime }a^{\prime }}(E)\;$is
calculated by ``attaching''\ the semi-infinite leads to the molecule\cite
{GF99}. As a result of the attachment the molecular levels acquire
a width that strongly depends on the coupling between
electrode and the molecule, $\Gamma \sim t_{{\rm Au-S}}^{2}/D_{{\rm Au}}$.
Here $D_{{\rm Au}}$ is the width of the s-band for Au electrodes, $t_{{\rm %
Au-S}}$ is determined mainly by the $sp\sigma $ hopping\ integral from Au to
the end sulfur atom on the BDT molecule, which is of the order of 1-2 eV.
One should expect significant broadening of the molecular levels 
when the molecule is attached by a thiol group to Au, since a strong
chemical bond is formed.

Under {\em zero} bias voltage, the electron charge $q_{m}$ on the site $m$
can be found from the Green's function in the standard manner as
\begin{eqnarray}
q_{m} &=&\sum_{a}\int_{-\infty }^{\infty }dEN_{ma}(E)f(E),  \label{eq:qm} \\
N_{ma}(E) &=&-\frac{1}{\pi }{\rm Im}G_{mama}(E),  \label{eq:Nma}
\end{eqnarray}
where $N_{ma}(E)$ is the density of states $a$ on the site $m$,
$f(E)=\left( 1+\exp \frac{E-E_{F}}{T}\right) ^{-1}$ is the Fermi function,
and $E_{F}$ is the chemical potential found from the global charge
neutrality of the system.

In the case of finite bias voltage the system is out of equilibrium and one
has to find the charge that is  ``flowing-in'' from the electrodes
onto the molecule, 
cf. \cite{Kirczenow00}. Then the DOS on the site $m,$ related to the influx
of electrons from the lead $w=1,2...,$ is written as
$N_{m}^{w}(E)=2({\rm for \ %
spin})\sum_{k_{z}^{w}{\bf k}_{||}^{w}}\left| \psi _{m}\left( k_{z}^{w},{\bf k%
}_{||}^{w}\right) \right| ^{2}\delta \left( E-E_{k_{z}{\bf k}%
_{||}}^{w}\right),$ where $\psi _{m}^{w}\left( k_{z},{\bf k}_{||}\right) $
is the wave function at the molecular site $m$, which asymptotically 
becomes an incident Bloch wave in the lead $w$ far from the molecule
with the wave vectors $\left( k_{z},{\bf k}_{||}\right)$.
Now we can find the occupation number for that site on the molecule due
to charge flowing in from the lead $w$ as $q_{m}=\sum_{w}\int_{-\infty
}^{\infty }dE N_{m}^{w}\left( E\right) f_{w}(E),$ where $f_{w}(E)$\ is the
Fermi function for the $w$th lead (i.e. with $E_{F}=E_{Fw}$). 
In order to calculate the Green's function (and the charges $q_m$) we
define the ``channels'' such that $k_{z}=k_{zl}^{w}(E),$ where $%
l=1,M $ enumerates all the quantum states in the lead unit cell (slice) \cite
{GF99}. 
It is convenient to re-write the expression for the charges in
terms of ``open channels''. The ``open channel'' is defined as a Bloch
wave which propagates in the lead at a given energy. 
The Bloch waves incident on the molecule 
(i.e. having the velocities towards the scatterer, $v_{l}>0$)
will contribute to the charge flowing to the molecule from a particular lead:
\begin{eqnarray}
N_{m}^{w}(E) &=&2\sum_{{\bf k}_{||}^{r}}\frac{1}{2\pi }\int_{-\pi
/d_{z}^{w}}^{\pi /d_{z}^{w}}dk_{z}^{r}\left| \psi _{m}\left( k_{z}^{w},{\bf k%
}_{||}^{w}\right) \right| ^{2}\delta \left( E-E_{k_{z}{\bf k}%
_{||}}^{w}\right)  \nonumber \\
&=&2\frac{1}{2\pi }\sum_{{\bf k}_{||}^{w}}\sum_{l\left( v_{l}^{w}>0\right) }%
\frac{1}{\hbar v_{l}^{w}}\left| \psi _{m}^{w}\left( k_{zl},{\bf k}%
_{||}\right) \right| ^{2},  \label{eq:nmch}
\end{eqnarray}
where $d_{z}$ is the unit cell length along the lead. $\psi ^w_m$ 
are normalized for the length of the wire, which drops out of the
final answers.
Note that the integration in (\ref{eq:nmch})\
goes over the {\em whole} Brillouin zone, not just over $k_{z}>0$. 
The delta-function picks up the open channels on the leads.
{}From now on we can drop the lead index and assume that one can later sum up
all the charges flowing from all the leads. Once the Hamiltonian is set up,
one calculates the charges on the sites, recalculates the onsite energies $%
\epsilon _{ma}$ and continues iteratively until the charges converge.

The current through the film is given by a standard expression \cite
{Landauer,GF99} 
\begin{equation}
I=\frac{2q}{h}\int dE\left[ f\left( E-\frac{qV}{2}\right) -f\left( E+\frac{qV%
}{2}\right) \right] T(E),  \label{eq:tok}
\end{equation}
where $q=|e|$ is the elementary charge, and $T(E)$ is the transmission probability
\begin{equation}
T(E)\equiv \sum_{{\bf k}_{\parallel },nn^{\prime }}\left| t_{nn^{\prime }}(E,%
{\bf k}_{\parallel })\right| ^{2},
\end{equation}
where the summation goes over the surface Brillouin zone of the lead.
Transmission coefficients $t_{nn^{\prime }}(E,{\bf k}_{\parallel })$\
between the scattering channels $n$ and $n^{\prime }$ are found from
the solution of the scattering problem\cite{GF99}.
In the case of weak molecule-electrode bonding
the transmission probability is approximately given by the Breit-Wigner
formula \cite{KB01}
\begin{equation}
T(E)\approx \sum_{r}\frac{\Gamma _{rL}\Gamma _{rR}}{\left( E-E_{r}\right)
^{2}+(\Gamma _{rL}+\Gamma _{rR})^{2}/4},  \label{eq:BW}
\end{equation}
where $E_{r}$ enumerates the energies of the molecular orbitals (MOs)
contributing to transport (not all of them do, see
Figs.~\ref{fig:TENE}, \ref{fig:TEgate}), $\Gamma _{rL(R)}/\hbar$ is the rate 
of the carrier transfer to the left (right) electrode from the
molecular orbital $r$. This formula applies
when the width of the MOs is much smaller than the energy
difference between them, so that the resonances do not overlap. Each
conducting molecular orbital produces a step-like
contribution to the current. Indeed, when the resonance falls into the
``window'' between the lowest and the 
highest Fermi levels in the leads $E_{FL}<E_{r}<E_{FR}$,
the current obtained from Eq.~\ref{eq:tok} is
\begin{equation}
I\approx \frac{2q}{\hbar }\frac{\Gamma _{rL}\Gamma _{rR}}{\Gamma _{rL}+\Gamma
_{rR}}.  \label{eq:Ires}
\end{equation}

It follows from this analysis that the current-voltage characteristic should
look as a series of steps, occurring when the resonant conditions are
satisfied for particular conducting molecular orbital. The apparent negative
differential resistance (NDR) at bias above 2V,
Figs.~\ref{fig:IVtop}-\ref{fig:spd}, results not from 
resonant tunneling but from the electrode density of states. In the present
model the electrode DOS in bounded from above for each particular value of $%
{\bf k}_{\Vert }.$ As a result, the current will be zero at the energies
above some threshold. The apparent NDR\ persists in the present calculations
irrespectively of the number of basis functions (s-, sp-, or spd-basis),
Fig.~\ref{fig:spd}.

\section{Effects of contact geometry and gating on current-voltage
characteristics}

We argue below that gating of SAMs in the experiments
\cite{SchonBDT01,SchonNi02} may have led to changes in
molecule-electrode geometry and, consequently, to large changes in conductance.
Given the strong orientational dependence of the current through conjugated
molecules like BDT, and that in experimental SAMs the molecules are
never positioned 
strictly normal to the electrode surface (as was assumed in
Refs.\cite{maxL00,maxA00}), we shall present the results for the transmission,
density of states (Figs.~\ref{fig:TENE}, \ref{fig:TEgate}) and I-V
curves (Figs.~\ref{fig:IVtop}-\ref{fig:spd}) for a series of 
tilting angles $\theta $ between the backbone of the molecule and the normal to the
Au(111) surface. The $\theta -$dependence of the I-V curves for BDT on the
top site and hollow site is illustrated on
Figs.~\ref{fig:IVtop},\ref{fig:IVhol} for $\theta 
=0-30^{\circ }$. It is especially strong for BDT on the top site. The
majority of the results is given for $\theta =10^{\circ },$ which seems to
be a reasonable choice for experimental SAMs. Note that in the upright
position $\theta =0$ in top site (i.e. perpendicular to the contact surface,
as was assumed in Ref.~\cite{maxL00}) the overlap between the S $x$ and $y$
p-orbitals ($xy$ being in contact surface plane, $z$ normal to the contact)
and the s-orbital on Au (or jellium) is {\em exactly zero} by symmetry,
since $(x|H|s)=(y|H|s)\equiv 0,$ where $H$ is the Hamiltonian. The
s-electron on the top Au can only hop onto a S $z$-orbital via a $%
(z|H|z)=ss\sigma $ hopping integral. Obviously, for BDT on the top site,
this result holds for all incident electrons with any ${\bf k}_{\parallel }.$
Therefore, the $x$ and $y$ p-orbitals on S cannot be traversed by electrons
incident from the contact. At the same time, only those states on the sulfur
ion are coupled to conjugated $\pi -$orbitals on the benzene ring.
Therefore, for the BDT on the top site and oriented normal to contact, the
current will be {\em suppressed}, as observed in  calculations by Di Ventra 
{\it et al}.\cite{maxL00,maxA00}. Obviously, this symmetry selection rule is
lifted for any $\theta \neq 0$. Thus, the previous calculations
\cite{maxL00,maxA00}  have
been performed at an artificial singular point. Incidentally, the same
conclusion applies to the scattering of the carriers incident with ${\bf k}%
_{\parallel }=0$ (surface $\Gamma -$point) on upright BDT on a hollow site.
Indeed, in this case the matrix element for hopping to the sulfur atom on
the molecule is proportional to $\sum_{i}(x_{i}|H|s)\exp (i{\bf k}%
_{\parallel }{\bf \rho }_{i})\propto \sum_{i}l_{i}=0$ for ${\bf k}%
_{\parallel }=0,$ where $l_{i}$ are the directional cosines connecting the
center of the triangle formed by three Au atoms on Au(111) surface with Au
atoms in the corners at positions ${\bf \rho }_{i}$. The same is obviously
true of the hopping to the $y$ p-orbital on S. Thus, in the case of BDT on a
hollow site, {\em all} the contribution to the current comes from states with $%
{\bf k}_{\parallel }\neq 0.$ Therefore, for BDT\ placed upright on the
hollow site, the total current is {\em not} suppressed, as it is for BDT on
the top site. Consequently, the current for BDT on the hollow site is
considerably less sensitive to the precise contact geometry.

\subsection{Density of states and transmission}

\begin{figure}[t]
\epsfxsize=8cm
\epsffile{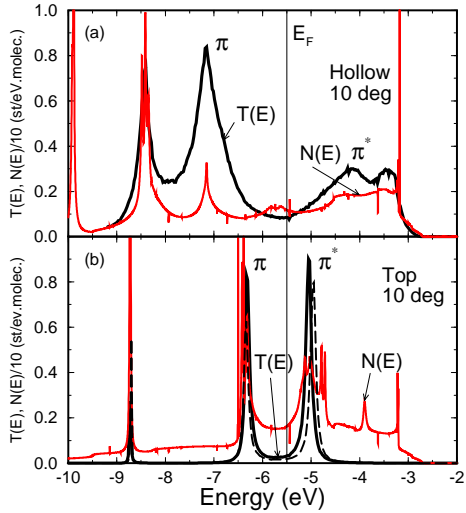}
\caption{ Density of states $N(E)$ and transmission $T(E)$ through
benzene-dithiolate (BDT) molecule on Au(111) as a function of energy: (a)
BDT on the hollow site, (b) BDT on the top site (see text for the
description of the configuration). 
The broken line indicates the transmission $T(E)$ under the bias
voltage 2V.
Molecules in both configuration are
tilted by 10$^o$. }
\label{fig:TENE}
\end{figure}

\begin{figure}[t]
\epsfxsize=8cm
\epsffile{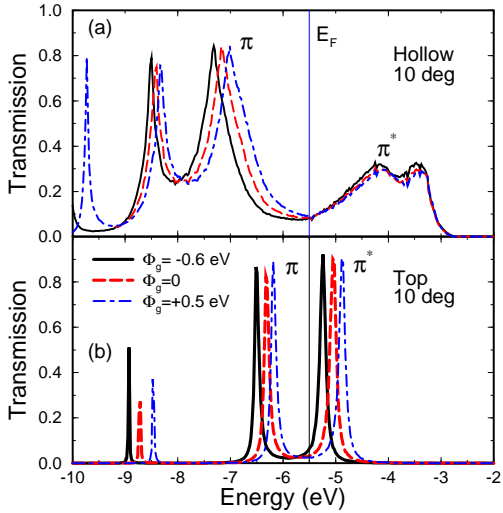}
\caption{ Transmission $T(E)$ through the BDT molecule
on Au(111) and the effect of gating: (a) hollow site, (b) top site. The gating is simulated by
shifting the on-site energies on BDT by $\Phi_g=$0.5, 0, and -0.5eV. Note
the presence of a sharp peak in $T(E)$ for top site (originating from the
LUMO on BDT) close to the Fermi level $E_F$. }
\label{fig:TEgate}
\end{figure}

Most of the present results can be appreciated from the analysis of the
transmission probability $T(E)$ and the density of states $N(E)\;$on the
BDT molecule, see Figs.~\ref{fig:TENE},\ref{fig:TEgate}. One expects
from the golden rule that the transmission 
would be proportional to the density of states. However, although peaks of
both functions follow each other rather closely, there are important
differences between the density of states and the
transmission. It is easier to analyze the results for the top position
first. There are two sharp peaks around the Fermi level $E_F $,
marked as $\pi ^{\ast }$ (at $E_{\pi ^{\ast }}=E_{F}+0.5$ eV) and $\pi $ (at 
$E_{\pi }=E_{F}-1.0$ eV), Fig.~\ref{fig:TEgate}(a). Transmission is
almost zero at $E>E_{\pi ^{\ast }},$ but there is a large density
of {\em non-conducting
} states in this energy interval. Those non-conducting states are formed at the end
of the molecule and reside primarily on the end sulfur atom and gold atom on
top of which the molecule sits, with little coupling to C$\pi $ conducting
states on the ring. The $\pi ^{\ast }$ peak contains mostly Au and S states
and some C$\pi $ ring states whereas the $\pi $ peak is made mostly of S and
C$\pi $ ring states with a little addition of Au states. Sulfur atoms
introduce the states in the HOMO-LUMO gap of the benzene ring (which is
about $6.5$ eV) and hybridize with C$\pi $ states to make the conducting
pathways across the BDT molecule. As a result, we see a much smaller gap
between $\pi $ and $\pi ^{\ast }$ states in BDT, which is $E_{\pi ^{\ast
}}-E_{\pi }=1.2$ eV. In the case of the hollow-site position, the situation
is considerably different. There the $\pi $ and $\pi ^{\ast }$ states are
much broader and pushed apart by much stronger hybridization with three
underlying Au atoms than for the top site,
Fig.~\ref{fig:TEgate}(b). The ``soft'' energy 
gap for the hollow position is $E_{\pi ^{\ast }}-E_{\pi }=3.25$ eV,
Fig.~\ref{fig:TEgate} (a) (cf.\cite{Kirczenow98}).

It is instructive to compare the present results with the recent data
for Ni-BDT-Ni SAMs \cite{SchonNi02} and jellium-LDA
calculations\cite{maxL00}. The systems should have many 
similarities since Au and Ni have almost the same work function of
about 5.1 eV\cite{Sze}. In both systems a large conductance
peak  is observed at a bias of about $V=2.1$V, which is larger than the
earlier value of 1.4V \cite{Reed97}. The peak in Ni system is narrower
and does not show any appreciable spin splitting. 
Interestingly, in Ni-BDT-Ni 
the additional  resonant-like features are found at low bias voltages of
about 0.3 V and 0.9 V. They correspond to smaller conductance compared
to the peak at 2.1V.
The position of the peaks should be compared with that of Ni
d-states. It is well known that the energy of a  minority peak in Ni
DOS is very close to the Fermi level, whereas the majority peak is
about 0.5eV below $E_F$. If the conducting molecular states
were considerably smeared out, like in the case of hollow position,
Fig.~\ref{fig:TENE}, then the observed low-bias features might be due
to those peaks in Ni d-DOS. The fact that the first peak is observed
at 0.3V, and not much smaller bias, may be due to energy dependence of
the molecular density of ``tail'' states at $E_F$, which shifts the
peaks by 0.3V. 
If, on the other hand, the peaks in the molecular
density of states are sharp, like in the case of top position,
Fig.~\ref{fig:TENE}, then  the peaks in conductance should correlate
with the position of the molecular orbital closest to the Fermi level (LUMO,
according to the present work). The position of the LUMO in BDT in the
present model is about
0.5 eV above the Fermi level for the top position.
Therefore, we expect that in the top-site
configuration there should be two peaks in the conductance, one at about
0.5-1.0 V (the position of the d-peak in Ni DOS with respect to the $\pi
^{\ast }$ resonance, depending on the connection between the molecule and
the electrodes) and another at about 0.5 V higher than the first one (at
1.0-1.5 V). Interestingly, this is very similar to what was observed
experimentally for the Ni-BDT-Ni system, with the peaks at $V=0.3$ V and $0.9
$ V \cite{SchonNi02}. 
If, on the other hand, the LUMO ($\pi^*$ state) is at 1eV above the
Fermi level, than the spin-peaks in Ni d-DOS might have produced the
peaks in conductance at 2-3V. Indeed, there is a conductance peak at
2.1V in both Ni- and Au-based systems, but it is not spin-split in the
case of Ni electrodes. It is worth mentioning that the position of the
first peak in conductance is a strong function  of the tilt angle,
and it may vary significantly, Fig.~\ref{fig:IVtop}.
Finally, we note that the HOMO-LUMO gap 
 in the ``jellium'' calculations is $\sim 5$ eV
\cite{maxL00}, which is substantially larger than that in bare BDT
molecules, and that is unlikely. The calculated value of the first peak in
conductance in jellium-LDA is 2.4V, larger than the observed value of
2.1V. This suggests that the calculated gap is {\em larger} than the
measured one, which is {\em contrary} to the usual notion that the LDA
gaps are smaller than the experimental ones due to insufficient
account of electron-electron correlations. 
As follows from this discussion, one needs more analysis to draw
definitive conclusions about the position of the lowest conducting
orbital with respect to the Fermi level in electrodes in Au-BDT-Au and
Ni-BDT-Ni.

\subsection{Gating the molecules}

\begin{figure}[t]
\epsfxsize=8cm
\epsffile{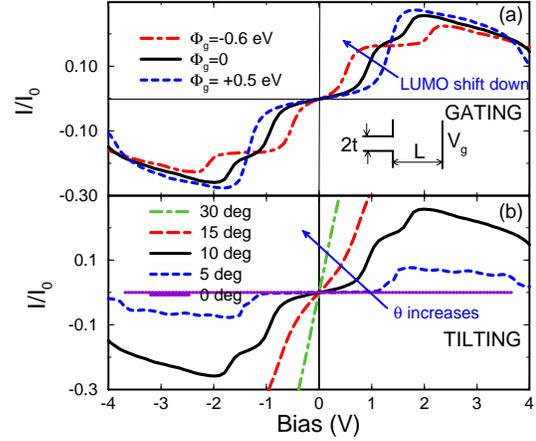}
\caption{ Current-voltage characteristic and effects of gating and
tilting with respect to the Au(111) electrode 
surface on current through the BDT molecule on the top site: (a) effect of
the gating, onsite energies are shifted by the amount $\Phi_g$
indicated on the figure, (b) effect of increasing tilt angle
$\theta$. Current is in units of $I_0=77.5\protect\mu A$. Inset:
schematic of the gate geometry, $2t=$1 nm, $L=4-5$ nm (nominal oxide
thickness was 30nm), $V_g$ is the
gate voltage.}
\label{fig:IVtop}
\end{figure}

The {\em gating effect} on the transmission and I-V\ characteristics is
illustrated in Figs.~\ref{fig:IVtop},\ref{fig:IVhol}. The gating is
modeled by shifting the on-site 
energies on the molecule by $\Phi _{g}$, which is usually $-0.5$, $0$,
and $0.5$ eV in the calculations. Obviously, 
in the experimental situation \cite{SchonBDT01} such a large shift would
require very large gating fields, comparable to the atomic fields in the
order of magnitude. This is because one has to substantially change the
electronic states on the molecule, and the characteristic energy is given by
the HOMO-LUMO\ gap, usually a few electron-volts. Such large fields
could not be possibly produced in the slot geometry with the channel length
of only $t=1-2$ nm through the gate oxide with thickness $L=4-5$ nm \cite
{SchonBDT01}. 
Schematic  of this gate is shown in inset in Fig.~\ref{fig:IVtop}(a). 
The analytical solution to this electrostatic problem can be
found by standard methods and it naturally contains a small parameter 
$t/L\ll 1,$ so the gating on the molecule itself would be much smaller than
the nominal gating voltage $V_{g}.$ 

One can speculate that large gating may
result from e.g.  charge accumulation in the gate oxide next to the
molecular film. However, changing the oxide from SiO$_{2}$ to Al$_{2}$O$_{3}$
apparently has not modified the results\cite{SchonBDT01}. Besides,
there is an abrupt change of 
conductance by about an order of magnitude at the gate voltage $%
V_{g}=-0.3$V, which would suggest a high sensitivity of the interface charge
to the bias voltage. Both facts are difficult to reconcile with the idea of
interface charge accumulation but we will study this possibility.

Comparing Figs.~\ref{fig:IVtop}(a) and \ref{fig:IVhol}(b), one observes
that the smaller hybridization 
between S p- and Au s-states for BDT on the top site produces sharp features
in the energy dependence of the transmission at the Fermi level. The
LUMO in this case is 
above the Fermi level by only about 0.5eV. The shifts of on-site energy
by similar amount substantially change the transmission at the Fermi
level, Fig.~\ref{fig:IVtop}(a), and the corresponding current per molecule
as shown in Fig.~\ref{fig:IVhol}(a). There
is a pseudogap at low voltages $V\lesssim 1$V, with the threshold voltage
moving by an amount comparable to the external shift $\Phi _{g}$ for the
top-site configuration. By contrast, the large hybridization of S p-states
with Au on the hollow site results in much broader energy tails of the
resonant peaks in the density of states in the gap region in the vicinity of
the Fermi level. Consequently, gating effect on the transmission,
Fig.~\ref{fig:IVhol}(b), and current, Fig.~\ref{fig:IVhol}(b), 
is smaller compared to the top-site situation. 

For the hollow-site
configuration there is no trace of the HOMO-LUMO gap in the I-V curve,
and the I-V curve is almost {\em ohmic} in the wide range of
voltages $V<2$V (Fig.~\ref{fig:IVhol}). This is indicative of the {\em
metalization} of 
the chemically bonded molecule.  This should have general implications, the
simplest being an obvious difficulty in gating such molecules. This
relates well to the large observed values of conductance (close to a
conductance quantum) through a single molecule, and its small
temperature dependence \cite{bao}.

Finally, it
is important to mention that in the present as well as other
calculations, the LUMO 
is the closest molecular orbital to the Fermi level $E_{F}$ and the maximum
gating effect is naturally expected when its energy is pulled down closer to 
$E_{F}.$ This takes place at {\em positive} gating voltage, and not the negative
one, as observed experimentally\cite{SchonBDT01}. This is an apparent
contradiction which needs to be resolved.

\subsection{Effect of contact geometry}

\begin{figure}[t]
\epsfxsize=8cm
\epsffile{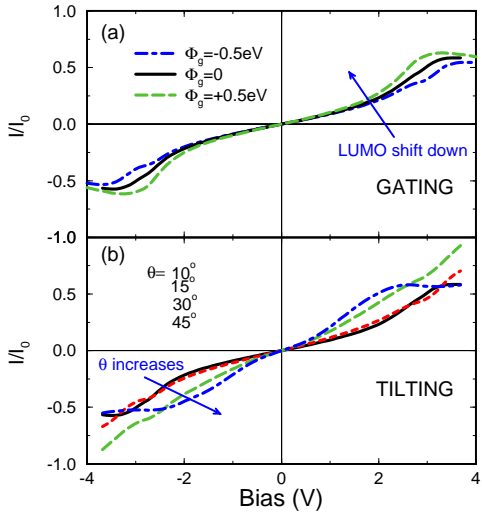}
\caption{Current-voltage characteristic and effects of gating and
tilting with respect to the Au(111) electrode 
surface on current through the BDT molecule on the hollow site: (a) effect
of the gating, (b) effect of increasing tilt angle. Current is in units of $%
I_0=77.5\protect\mu A$. }
\label{fig:IVhol}
\end{figure}

The tilting angle has a large effect on the I-V curves of BDT
molecules, see
Figs.~\ref{fig:IVtop}, \ref{fig:IVhol}.
The behavior of the BDT on 
the top site and on the hollow site is again rather different. 
The I-V curve for the hollow site
remains ohmic for tilting angles up to 75$^{\circ }$ with moderate changes
of conductance, Fig.~\ref{fig:IVhol}(b). 
The variation of the current with the angle $\theta$
are much larger for the top site, Fig.~\ref{fig:IVhol}(a). By changing
$\theta $ 
from 5$^{\circ } $ to just 15$^{\circ }$, one drives the I-V\ characteristic
from one with a gap of about 2V to the ohmic one with a large relative
change of conductance. Even changing $\theta$ from 10$^{\circ }$ to 15$%
^{\circ }$ changes the conductance by about an order of magnitude. 

\begin{figure}[t]
\epsfxsize=7.5cm
\epsffile{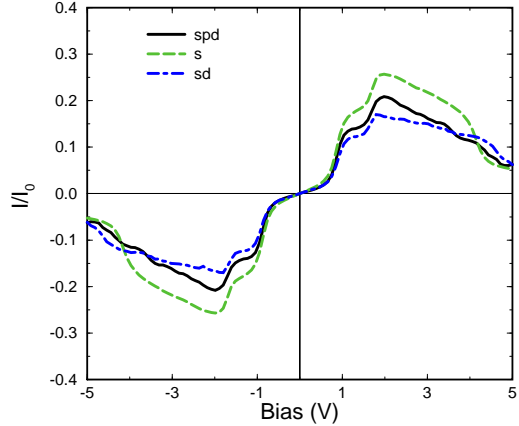}
\caption{ Effect of electrode structure of Au(111) electrodes on current
through the BDT molecule on the top site. There is moderate difference
between calculations using s-, sp-, and full spd-bases. Current is in units
of $I_0=77.5\protect\mu A$.}
\label{fig:spd}
\end{figure}

Finally, Fig.~\ref{fig:spd} illustrates the role of the electronic structure of the
electrodes. The results described above have been obtained with only the
s-states on Au atoms. We have also considered an sp- and spd-basis for Au.
Although substantially increases the computing time,
but the addition of p- and
d-states brings about only moderate changes in current. Since the
hybridization is different for different cases, the current magnitude
slightly varies for different basis sets.

\subsection{Possible origins of the observed gating effect}

With regards to the origin of the observed gating effect, one can envisage
that in dilute BDT-alkane solution sandwiched between Au electrodes there
may be two processes going on that significantly change conductance. Since
the BDT\ molecules are clamped by the matrix of alkane chains, they have to
move with it. Note that the matrix of alkane chains is not in registry with
the Au electrode. The BDT molecules have a nominal length of 7.2\AA and are
dissolved in the matrix of (CH$_{2}$)$_{5}$S alkanethiol insulating
molecules with a nominal length of 8.3\AA . Thus, the BDT molecule would
appear as a dip on the surface of the matrix, and one or a few gold atoms
can get into this dip during the deposition and bind to the end S. The
geometry of this bond is uncertain, and the bond may well be stretched. In this
case even a slight perturbation exerted on the SAM might lead to
reconfiguration of the bond resulting in large changes of conductance.
 It seems reasonable to assume that the ``domain
walls'' separating different patches in an alkane matrix move rather
freely in the system, since\ it does not require much energy. The BDT
molecules will follow the matrix and can either snap from a hollow site to a
top site and back and/or change the tilt angle. Both processes may be
accompanied by large changes in the conductance. Conformational changes of
the clamped BDT molecule are rather restricted, and the motion of the
``domain walls'' may be quite repeatable. One may wonder what causes the
domain walls to move. As a possible reason, we suggest the presence of 
positive metal ions inside the organic
film, as a small concentration of electrode ions in a SAM is 
rather inevitable. Indeed, Au$^{+}$ strongly interacts with C$_{6}$H$_{6}$
in the gas phase and forms an Au$^{+}-$C$_{6}$H$_{6}$ complex with a binding
energy 2.65 eV, whereas neutral Au forms a Van-der-Waals complex with the
binding energy 90 meV\cite{Aubenzene}. Even more likely is a formation
of those complexes with thiophenes, which carry an electric dipole. 
It is likely that a similar charged
complex can form with BDT molecules in a SAM with
those BDT molecules that 
have lost or are in poor contact with the gold substrate. 
It is also possible that a charged complex can be formed between the charged
metallic ions (Au$^+$ or other electrode metals) and alkane chains.  
A small field in
the organic film will then produce a tangential force on the ions, and this
may trigger the domain wall motion when the pinning is weak. Additionally,
since the packing of the film is not ideal (an organic film is usually a
rather disordered patchwork of ``grains''), the Maxwell force acting on the
top Au electrode at finite drain voltage and/or electron wind force may
trigger the domain wall motion, which may also require a combination of
these factors.
The second possibility would be a build-up of the interface charge, but
apparently the replacement of oxide did not change the results. As mentioned
above, it is also difficult to explain the jump in conductance at a certain
value of the gate voltage.

\section{Summary}

We have presented an extensive analysis of the electronic states and
transport through the benzene-dithiolate molecule, which is the simplest
conjugated molecule that forms a SAM and exhibits large changes of
conductance under gating\cite{SchonBDT01}. It shows that the effect of
gating strongly depends on the geometry of the molecule-electrode contact,
and is maximal for the less coordinated top-site position. It is related
to the sharpness of the peak in transmission, which corresponds to the LUMO
on BDT and is close to the Fermi level $E_{F}$ of an Au electrode. By the
same token, the current is more sensitive to the tilting angle of the
molecule when it is positioned on the less
coordinated top site. It is worth mentioning that the fact that the LUMO is
closer to $E_{F}$ suggests that a {\em positive} gating voltage $V_{g}$
should produce the larger effect, not a negative gate voltage, as was
observed in \cite{SchonBDT01}. 
This discrepancy should be addressed in
the future. Binding on the highly coordinated hollow site naturally
leads to large hybridization of the molecular states with electrode states,
which become smeared out.
Consequently, BDT molecule becomes {\em %
metallized}, i.e. the I-V characteristic becomes practically ohmic. A very small
effect of gating is predicted for this geometry. In any case, it is
difficult to explain the large observed effect of gating on BDT molecules in
a slot with a width of only 1 nm by small voltage applied to the gate 4-5 nm
away. One should assume that there is either (i) a build-up of interface
charge in the immediate vicinity of the slot opening, which is very
sensitive to the gate voltage; or (ii) small inhomogeneous electrostatic
forces and resulting stresses on the SAM\ result in the reconfiguration of
the film and, consequently, of a clamped inside BDT molecule with respect to
the gold contacts. Both of those mechanisms have problems of their own in
explaining the experimental observations, as discussed in the text. 
Changes in random charges in the film and/or chemical composition of
the BDT molecules (e.g. loss of end sulfur) are possible but are
unlikely to be reversible.  One
needs to characterize the films better and vary the gate oxide thickness and
other parameters of the system in order to confirm one of those mechanisms
or suggest some other effects controlling the gating in slot geometries like
the one used in Refs. \cite{SchonBDT01,SchonNi02}.

We thank J.H. Sch\"{o}n for extensive helpful conversations and sharing his
data and preprints, and R.S. Williams for useful discussions.
The work has been partly supported by DARPA.



\end{document}